\RequirePackage{lineno}
\documentclass[aps,prl,twocolumn,showpacs,superscriptaddress,letterpaper,amsmath,amssymb]{revtex4}

\usepackage{graphicx}  
\usepackage{dcolumn}   
\usepackage{bm}        
\usepackage{amssymb}   
\usepackage{feynmf}
\usepackage{slashed}
\usepackage{subfig}
\def\gsim{\lower0.5ex\hbox{$\:\buildrel >\over\sim\:$}}
\def\lsim{\lower0.5ex\hbox{$\:\buildrel <\over\sim\:$}}

\def \n{\noindent}

\let\n=\nu
\let\t=\tau

\newcommand{\be}{\begin{equation}}
\newcommand{\ee}{\end{equation}}
\newcommand{\bea}{\begin{eqnarray}}
\newcommand{\eea}{\end{eqnarray}}

\newcommand{\nbox}{{\,\lower0.9pt\vbox{\hrule \hbox{\vrule height 0.2 cm
\hskip 0.2 cm \vrule height 0.2 cm}\hrule}\,}}

\def\missET {{\not\!\! E_T}}

\begin{document}

\thispagestyle{empty}
\vspace*{-3.5cm}

\vspace{0.5in}


\bibliographystyle{apsrev}

\title{{Search for a heavy particle decaying to a \\top quark and a
    light quark in {\bf $p\bar{p}$} collisions at {\bf $\sqrt{s}=1.96$ TeV}}}

\affiliation{Institute of Physics, Academia Sinica, Taipei, Taiwan 11529, Republic of China}
\affiliation{Argonne National Laboratory, Argonne, Illinois 60439, USA}
\affiliation{University of Athens, 157 71 Athens, Greece}
\affiliation{Institut de Fisica d'Altes Energies, ICREA, Universitat Autonoma de Barcelona, E-08193, Bellaterra (Barcelona), Spain}
\affiliation{Baylor University, Waco, Texas 76798, USA}
\affiliation{Istituto Nazionale di Fisica Nucleare Bologna, $^{ee}$University of Bologna, I-40127 Bologna, Italy}
\affiliation{University of California, Davis, Davis, California 95616, USA}
\affiliation{University of California, Irvine, Irvine, California 92697, USA}
\affiliation{University of California, Los Angeles, Los Angeles, California 90024, USA}
\affiliation{Instituto de Fisica de Cantabria, CSIC-University of Cantabria, 39005 Santander, Spain}
\affiliation{Carnegie Mellon University, Pittsburgh, Pennsylvania 15213, USA}
\affiliation{Enrico Fermi Institute, University of Chicago, Chicago, Illinois 60637, USA}
\affiliation{Comenius University, 842 48 Bratislava, Slovakia; Institute of Experimental Physics, 040 01 Kosice, Slovakia}
\affiliation{Joint Institute for Nuclear Research, RU-141980 Dubna, Russia}
\affiliation{Duke University, Durham, North Carolina 27708, USA}
\affiliation{Fermi National Accelerator Laboratory, Batavia, Illinois 60510, USA}
\affiliation{University of Florida, Gainesville, Florida 32611, USA}
\affiliation{Laboratori Nazionali di Frascati, Istituto Nazionale di Fisica Nucleare, I-00044 Frascati, Italy}
\affiliation{University of Geneva, CH-1211 Geneva 4, Switzerland}
\affiliation{Glasgow University, Glasgow G12 8QQ, United Kingdom}
\affiliation{Harvard University, Cambridge, Massachusetts 02138, USA}
\affiliation{Division of High Energy Physics, Department of Physics, University of Helsinki and Helsinki Institute of Physics, FIN-00014, Helsinki, Finland}
\affiliation{University of Illinois, Urbana, Illinois 61801, USA}
\affiliation{The Johns Hopkins University, Baltimore, Maryland 21218, USA}
\affiliation{Institut f\"{u}r Experimentelle Kernphysik, Karlsruhe Institute of Technology, D-76131 Karlsruhe, Germany}
\affiliation{Center for High Energy Physics: Kyungpook National University, Daegu 702-701, Korea; Seoul National University, Seoul 151-742, Korea; Sungkyunkwan University, Suwon 440-746, Korea; Korea Institute of Science and Technology Information, Daejeon 305-806, Korea; Chonnam National University, Gwangju 500-757, Korea; Chonbuk National University, Jeonju 561-756, Korea}
\affiliation{Ernest Orlando Lawrence Berkeley National Laboratory, Berkeley, California 94720, USA}
\affiliation{University of Liverpool, Liverpool L69 7ZE, United Kingdom}
\affiliation{University College London, London WC1E 6BT, United Kingdom}
\affiliation{Centro de Investigaciones Energeticas Medioambientales y Tecnologicas, E-28040 Madrid, Spain}
\affiliation{Massachusetts Institute of Technology, Cambridge, Massachusetts 02139, USA}
\affiliation{Institute of Particle Physics: McGill University, Montr\'{e}al, Qu\'{e}bec, Canada H3A~2T8; Simon Fraser University, Burnaby, British Columbia, Canada V5A~1S6; University of Toronto, Toronto, Ontario, Canada M5S~1A7; and TRIUMF, Vancouver, British Columbia, Canada V6T~2A3}
\affiliation{University of Michigan, Ann Arbor, Michigan 48109, USA}
\affiliation{Michigan State University, East Lansing, Michigan 48824, USA}
\affiliation{Institution for Theoretical and Experimental Physics, ITEP, Moscow 117259, Russia}
\affiliation{University of New Mexico, Albuquerque, New Mexico 87131, USA}
\affiliation{The Ohio State University, Columbus, Ohio 43210, USA}
\affiliation{Okayama University, Okayama 700-8530, Japan}
\affiliation{Osaka City University, Osaka 588, Japan}
\affiliation{University of Oxford, Oxford OX1 3RH, United Kingdom}
\affiliation{Istituto Nazionale di Fisica Nucleare, Sezione di Padova-Trento, $^{ff}$University of Padova, I-35131 Padova, Italy}
\affiliation{University of Pennsylvania, Philadelphia, Pennsylvania 19104, USA}
\affiliation{Istituto Nazionale di Fisica Nucleare Pisa, $^{gg}$University of Pisa, $^{hh}$University of Siena and $^{ii}$Scuola Normale Superiore, I-56127 Pisa, Italy}
\affiliation{University of Pittsburgh, Pittsburgh, Pennsylvania 15260, USA}
\affiliation{Purdue University, West Lafayette, Indiana 47907, USA}
\affiliation{University of Rochester, Rochester, New York 14627, USA}
\affiliation{The Rockefeller University, New York, New York 10065, USA}
\affiliation{Istituto Nazionale di Fisica Nucleare, Sezione di Roma 1, $^{jj}$Sapienza Universit\`{a} di Roma, I-00185 Roma, Italy}
\affiliation{Rutgers University, Piscataway, New Jersey 08855, USA}
\affiliation{Texas A\&M University, College Station, Texas 77843, USA}
\affiliation{Istituto Nazionale di Fisica Nucleare Trieste/Udine, I-34100 Trieste, $^{kk}$University of Udine, I-33100 Udine, Italy}
\affiliation{University of Tsukuba, Tsukuba, Ibaraki 305, Japan}
\affiliation{Tufts University, Medford, Massachusetts 02155, USA}
\affiliation{University of Virginia, Charlottesville, Virginia 22906, USA}
\affiliation{Waseda University, Tokyo 169, Japan}
\affiliation{Wayne State University, Detroit, Michigan 48201, USA}
\affiliation{University of Wisconsin, Madison, Wisconsin 53706, USA}
\affiliation{Yale University, New Haven, Connecticut 06520, USA}

\author{T.~Aaltonen}
\affiliation{Division of High Energy Physics, Department of Physics,
University of Helsinki and Helsinki Institute of Physics, FIN-00014,
Helsinki, Finland}
\author{J.~Adelman}
\affiliation{Yale University, New Haven, Connecticut 06520, USA}
\author{B.~\'{A}lvarez~Gonz\'{a}lez$^z$}
\affiliation{Instituto de Fisica de Cantabria, CSIC-University of Cantabria, 39005 Santander, Spain}
\author{S.~Amerio}
\affiliation{Istituto Nazionale di Fisica Nucleare, Sezione di Padova-Trento, $^{ff}$University of Padova, I-35131 Padova, Italy}
\author{D.~Amidei}
\affiliation{University of Michigan, Ann Arbor, Michigan 48109, USA}
\author{A.~Anastassov$^x$}
\affiliation{Fermi National Accelerator Laboratory, Batavia, Illinois 60510, USA}
\author{A.~Annovi}
\affiliation{Laboratori Nazionali di Frascati, Istituto Nazionale di Fisica Nucleare, I-00044 Frascati, Italy}
\author{J.~Antos}
\affiliation{Comenius University, 842 48 Bratislava, Slovakia; Institute of Experimental Physics, 040 01 Kosice, Slovakia}
\author{G.~Apollinari}
\affiliation{Fermi National Accelerator Laboratory, Batavia, Illinois 60510, USA}
\author{J.A.~Appel}
\affiliation{Fermi National Accelerator Laboratory, Batavia, Illinois 60510, USA}
% ...

\collaboration{CDF Collaboration\footnote{With visitors from
$^a$Istituto Nazionale di Fisica Nucleare, Sezione di Cagliari, 09042 Monserrato (Cagliari), Italy,
$^b$University of CA Irvine, Irvine, CA 92697, USA,
$^c$University of CA Santa Barbara, Santa Barbara, CA 93106, USA,
$^d$University of CA Santa Cruz, Santa Cruz, CA 95064, USA,
$^e$Institute of Physics, Academy of Sciences of the Czech Republic, Czech Republic,
$^f$CERN, CH-1211 Geneva, Switzerland,
$^g$Cornell University, Ithaca, NY 14853, USA,
$^h$University of Cyprus, Nicosia CY-1678, Cyprus,
$^i$Office of Science, U.S. Department of Energy, Washington, DC 20585, USA,
$^j$University College Dublin, Dublin 4, Ireland,
$^k$ETH, 8092 Zurich, Switzerland,
$^l$University of Fukui, Fukui City, Fukui Prefecture, Japan 910-0017,
$^m$Universidad Iberoamericana, Mexico D.F., Mexico,
$^n$University of Iowa, Iowa City, IA 52242, USA,
$^o$Kinki University, Higashi-Osaka City, Japan 577-8502,
$^p$Kansas State University, Manhattan, KS 66506, USA,
$^q$Ewha Womans University, Seoul, 120-750, Korea,
$^r$University of Manchester, Manchester M13 9PL, United Kingdom,
$^s$Queen Mary, University of London, London, E1 4NS, United Kingdom,
$^t$University of Melbourne, Victoria 3010, Australia,
$^u$Muons, Inc., Batavia, IL 60510, USA,
$^v$Nagasaki Institute of Applied Science, Nagasaki, Japan,
$^w$National Research Nuclear University, Moscow, Russia,
$^x$Northwestern University, Evanston, IL 60208, USA,
$^y$University of Notre Dame, Notre Dame, IN 46556, USA,
$^z$Universidad de Oviedo, E-33007 Oviedo, Spain,
$^{aa}$CNRS-IN2P3, Paris, F-75205 France,
$^{bb}$Texas Tech University, Lubbock, TX 79609, USA,
$^{cc}$Universidad Tecnica Federico Santa Maria, 110v Valparaiso, Chile,
$^{dd}$Yarmouk University, Irbid 211-63, Jordan,
}}
\noaffiliation

\begin{abstract}
 We present a search for a new heavy particle $M$ produced in
 association with a top quark, \mbox{$p\bar{p}\rightarrow t(M \rightarrow
 \bar{t}q)$} or \mbox{$p\bar{p}\rightarrow \bar{t}(\bar{M}\rightarrow t\bar{q})$}, where $q$ stands for up quarks
 and down quarks. Such a particle may explain the recent
 anomalous measurements of top-quark forward-backward asymmetry.  If the light-flavor quark ($q$) is
 reconstructed as a jet ($j$), this gives a
 $\bar{t}+j$ or $t+j$  resonance  in $t\bar{t}$+jet events, a previously
 unexplored experimental signature. In a
 sample of events with exactly one lepton, missing
 transverse momentum and at least five jets, corresponding to an integrated
 luminosity of 8.7 fb$^{-1}$ collected by the CDF II detector, we find the data to be consistent with the 
 standard model. We set cross-section upper limits on the production
 ($p\bar{p}\rightarrow Mt$ or $\bar{M}\bar{t}$) at 95\% confidence level 
  from $0.61$ pb to $0.02$ pb
 for $M$ masses ranging from $200$ GeV$/c^2$ to $800$ GeV$/c^2$, respectively.
\end{abstract}



\maketitle

\date{\today}

The standard model (SM) of particle physics has been extensively tested at
the Tevatron collider and in initial results from the Large Hadron
Collider.  A deviation from predictions may
provide a clue that reveals the solution to outstanding theoretical
concerns with the SM, such as the hierarchy problem or the
fermionic mass hierarchy.  One area of particular interest is
properties of the top quark, whose large mass suggests that it may
play a special role in electroweak symmetry breaking~\cite{topewsb}.
The Tevatron experiments
have performed detailed studies of the properties of the top quark.  Recently, CDF reported a
measurement of the top-quark production forward-backward asymmetry
($A_{fb}$) that is significantly larger than predicted by the
SM~\cite{cdfafb}, and is especially significant at large mass of the
$t\bar{t}$ system;
D0 reported~\cite{d0afb} a result consistent with the inclusive CDF
measurement, but without a high-mass enhancement. Many models have been built to explain such a discrepancy,
most involving the production of a new heavy mediating particle $M$
that enhances $A_{fb}$.  Many of these models~\cite{zurek2}
predict significant enhancements of the $t\bar{t}$ production
cross section~\cite{atlastop}, the single-top production cross section~\cite{atlasst}, or the
same-sign top-quark pair-production cross section~\cite{cmsss,atlasss}, none of which have
been confirmed in experimental tests.

One class of models~\cite{zurek,schwarz} evades the same-sign
top-quark limits by prohibiting the particle from acting as its own
antiparticle, and can satisfy the $t\bar{t}$ and single top-quark
cross-section constraints for some coupling values. In addition,
models in this class predict a new, unexplored
experimental signature: the production of a heavy new particle $M$  in
 association with a top quark ($p\bar{p}\rightarrow Mt$ or $p\bar{p}\rightarrow \bar{M}\bar{t}$) which decays
 via $M\rightarrow \bar{t}q$ or $\bar{M}\rightarrow t\bar{q}$. Since the light-flavor up quark
 or down quark ($q$) is
 reconstructed as a jet ($j$), the final state is $t\bar{t}+j$ with a resonance in the
 $\bar{t}+j$ or $t+j$ system, which has not
 been previously examined.  This Letter reports the first search for
 such a resonance.

We consider  the production mode $p\bar{p}\rightarrow
Mt \rightarrow t\bar{t}q \rightarrow W^{+}bW^{-}\bar{b}q$ (and its
conjugate $t\bar{t}\bar{q}$ mode) in which one $W$ boson
decays leptonically (including leptonic $\tau$ decays) and the
second $W$ boson
decays  to a quark-antiquark pair. This decay mode features large
branching ratios while reducing  to a manageable level the backgrounds
other than SM $t\bar{t}$ production. Such a signal is similar to SM top-quark
pair production and decay, but with an additional jet coming from the $M$
resonance decay.

We analyze a sample of events corresponding to an integrated
luminosity of 8.7$\pm 0.5$ fb$^{-1}$ recorded by the CDF II detector~\cite{cdf}, a general purpose detector designed to study collisions at the Fermilab Tevatron
$p\overline{p}$ collider at $\sqrt{s}=1.96$ TeV. CDF has a charged-particle tracking system consisting of a silicon microstrip tracker and a drift chamber that are immersed in a 1.4 T magnetic field~\cite{trackingsystems}. Electromagnetic and hadronic calorimeters surrounding
the tracking system measure particle energies, and an additional system of drift chambers
located outside the calorimeters detects muons.

Events enter this sample by satisfying  online selection criteria
(trigger), requiring an $e$ or $\mu$ candidate~\cite{leptonid} with
transverse momentum $p_T$~\cite{coordinates} greater than $18$ GeV$/c$. After trigger selection, events are retained if the electron or muon candidate has a pseudorapidity $|\eta| < 1.1$~\cite{coordinates}, $p_T > 20$ GeV$/c$ and satisfies the standard CDF identification and
isolation requirements~\cite{leptonid}. We reconstruct jets in the
calorimeter using the {\sc jetclu}~\cite{jetclu} algorithm with a
clustering radius of 0.4 in $\eta-\phi$ space, and calibrated using the techniques outlined in \cite{jes}. Jets are
selected if they have transverse energy $E_T > 15$ GeV and $|\eta| < 2.4$. Missing transverse momentum~\cite{metdef}
is reconstructed using fully corrected
calorimeter and muon information~\cite{leptonid}.

The signature of $t\bar{t}q \rightarrow W^{+}bW^{-}\bar{b}q
\rightarrow  \ell\nu bq\bar{q}'\bar{b}q$ (and the conjugate 
$t\bar{t}\bar{q}$ mode) is a charged lepton ($e$ or $\mu$), missing
transverse momentum, two jets from $b$-quarks and three jets from
light quarks.  We select events with exactly one electron or
muon, at least five jets, and missing transverse momentum greater than
20~GeV$/c$. Since a signal would have two jets with $b$-quarks, we require (with minimal loss of efficiency) evidence of decay of a $b$-hadron in at least one jet. This requirement, called $b$-tagging, makes use of the {\sc secvtx} algorithm~\cite{secvtx}.

We model the production of  $Mt$ and $\bar{M}\bar{t}$ with $m_M=200-800$
GeV$/c^2$ and subsequent decays  with {\sc madgraph}~\cite{madgraph}. Additional radiation, hadronization and
showering are described by {\sc pythia}~\cite{pythia}. The detector
response for all simulated samples is modeled by the {\sc geant}-based CDF II detector simulation~\cite{cdfsimul}.

The dominant SM background to the $t\bar{t}+j$ signature is top-quark
pair production with an additional jet due to initial-state or
final-state radiation. We model this background using
{\sc pythia} $t\bar{t}$ production with a top-quark mass $m_t = 172.5$ GeV/$c^2$, compatible with the best current determination~\cite{topmass}. We
normalize the $t\bar{t}$ background to the theoretical calculation at next-to-next-to-leading order (NNLO) in $\alpha_s$~\cite{nloa}.  In addition,
events generated by a next-to-leading order generator, {\sc mc@nlo}~\cite{mcatnlo} are also used in estimating an uncertainty
	    in modeling the radiation of an additional jet.

 The second largest SM background process is the associated production 
of a $W$ boson and jets. Samples of  $W$ boson + jets events with light- and heavy-flavor 
jets are generated using the {\sc alpgen}~\cite{alpgen} program,
interfaced with a parton-shower model 
from {\sc pythia}.  The $W$ boson + jets samples are normalized to the
measured $W$ boson production cross section, 
with an additional multiplicative factor for the relative contribution of heavy- 
and light-flavor jets, following the same technique utilized previously in measuring the
top-quark pair-production cross section~\cite{secvtx}. Multi-jet
background, in which a jet is
misreconstructed as a lepton, is modeled using a jet-triggered sample normalized
to a background-dominated region at low missing transverse momentum
where the multi-jet background is large.

The SM backgrounds due to single top quark and diboson production are modeled
using {\sc madgraph} interfaced with {\sc pythia} parton-shower models and {\sc pythia}, respectively, and normalized to next-to-leading-order
cross sections~\cite{ewknlo}. 

A signal may be observed as an excess of events above expectations
from backgrounds in event distributions versus the mass of the $tj$
system ($M\rightarrow tj$) or the $\bar{t}j$
system ($M\rightarrow \bar{t}j$). In \mbox{$t\bar{t}+j$} events,
 we first identify the jets belonging to the $t\bar{t}$ system, using a kinematic fitter~\cite{rad}
to select from all available jets in the event the four jets most consistent with the $t\bar{t}$ topology.  In the fit, the top-quark and $W$-boson
masses are constrained to be $172.5$ GeV/$c^2$ and $80.2$ GeV/$c^2$,
respectively. All remaining
jets are considered candidates for the light-quark jet in the $tj$ or  $\bar{t}j$
resonance. These remaining  jets each
are paired with the reconstructed top quark and anti-top quark, and the largest invariant mass of all
such pairings is chosen as the resonance-mass reconstruction,
$m_{tj}$. Backgrounds, in which there is no resonance,
 give a broad and low distribution of $m_{tj}$, while a signal would be reconstructed near the resonance mass.

\begin{figure}[htp]
\subfloat[$W$ boson + jet control region: at least $5$ jets, exactly zero $b$-tags.]{\includegraphics[width=0.6\linewidth]{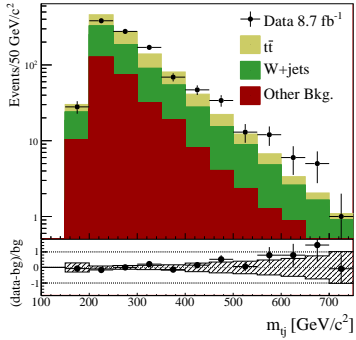}}\\
\subfloat[$t\bar{t}$ plus additional radiated jet control region: at
least $5$ jets, at least one $b$-tag, \mbox{$H_{T}<225$ GeV}.]{\includegraphics[width=0.6\linewidth]{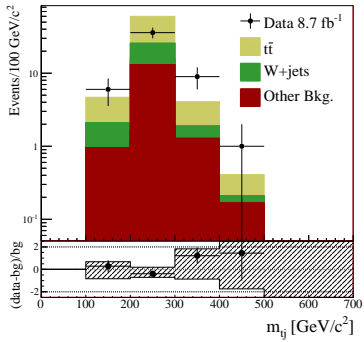}}\\
\caption{Distribution of events versus reconstructed $tj$ or $\bar{t}j$ invariant mass ($m_{tj}$) for observed data and expected backgrounds in two 
control regions. The lower panes give the relative
  difference between the observed and expected distributions; the
  hatched areas show the combined statistical and systematic
  uncertainties of the expected background.}
\label{fig:ctrl_region}
\end{figure}

\begin{figure}[!h]
\includegraphics[width=0.7\linewidth]{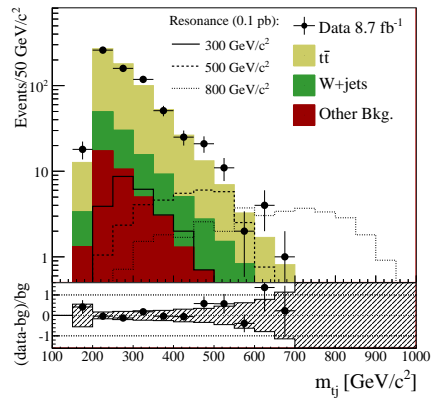}\\
\caption{Distribution of events versus reconstructed $tj$ or $\bar{t}j$ invariant mass,
  $m_{tj}$, for observed data and expected backgrounds in the signal region.  Three signal
  hypotheses are shown, assuming a total cross section of 0.1 pb.  The lower pane gives the relative
  difference between the observed and expected distributions; the
  hatched area shows the combined statistical and systematic
  uncertainties of the expected background.}
\label{fig:sig_region}
\end{figure}

We consider several sources of systematic uncertainty on the predicted background
rates and distributions, as well as on the expectations for a
signal. Each systematic uncertainty affects the expected sensitivity to new physics, expressed
as an expected cross-section upper limit in the no-signal
assumption. The dominant systematic uncertainty is the jet energy
scale (JES)~\cite{jes}, followed by theoretical uncertainties on the
cross sections of the background processes.  To probe the description
of the additional jet, we compare our nominal $t\bar{t}$ model to one
generated by {\sc mc@nlo} and take the full difference as a systematic
uncertainty.  We also consider systematic uncertainties associated with the description of
initial- and final-state radiation~\cite{rad}, uncertainties in the efficiency of
reconstructing leptons and identifying $b$-quark jets, and uncertainties in the contribution
from multiple proton interactions.  In addition, we consider
a variation of the $Q^2$ scale of $W$ boson+jet events in {\sc algpen}. In each case, we treat
the unknown underlying quantity as a nuisance parameter and measure the distortion of the
 $m_{tj}$ spectrum for positive and negative fluctuations of the
 underlying quantity. Uncertainties in the theoretical cross-section
 normalization are also included. Table~\ref{tab:syst} lists the
 contributions of each of these sources of systematic uncertainty.

\begin{table}[!h]
\caption{Contributions to systematic uncertainty on the two main
  expected background processes and the total background yield and
  from an example $500$ GeV$/c^2$ resonance signal with an assumed
  total cross  section of 0.1 pb.}
\begin{ruledtabular}
\begin{tabular}{lcccc}
 Process & $t\bar{t}$ & $W$ + jets &  Total Bg.& $Mt+\bar{M}\bar{t}$ \\ \hline
Yield & 550.6 & 78.6& 669.2 & 34.0\\  \hline
JES & $17\%$ & $15\%$ & $16\%$ & $9\%$ \\
Cross section & $10\%$ &  $30\%$  & $12\%$ & -\\
$t\bar{t}$ generator & $6\%$ &  -  & $5\%$ & -\\ 
Gluon radiation & $6\%$ & -& $5\%$  & $4\%$ \\
($e/\mu$, $b$-jet) ID eff.  & $5\%$ & $5\%$ & $5\%$ & $5\%$ \\
Mult. interactions & $3\%$ & $2\%$ & $3\%$ & $2\%$\\
$Q^2$ scale& - & $19\%$  &$2\%$  & - \\ \hline
Total syst. uncert.& $22\%$ &  $39\%$ & $22\%$ & $11\%$\\ 
\end{tabular}
\end{ruledtabular}
\label{tab:syst}
\end{table}

We validate our modeling of the SM backgrounds in three
background-dominated control regions. The $t\bar{t}$ background is
validated in events with exactly 4 jets and at least one $b$ tag. We
validate $W$ + jets backgrounds in events with at least 5 jets and no
$b$ tags. Finally, modeling of SM $t\bar{t}$ events with an additional
jet is validated by examining a signal-depleted region with at least 5
jets, at least one $b$ tag and $H_{T}$, the
scalar sum of lepton and jet transverse momenta, less than $225$ GeV. As shown in Fig.~\ref{fig:ctrl_region}, we find that the backgrounds are well modeled within systematic uncertainties.

\begin{figure}[!h]
\includegraphics[width=0.7\linewidth]{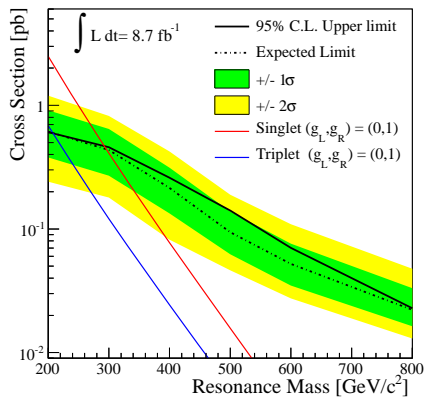}\\
\caption{Upper limits at 95\% CL on $t\bar{t}+j$ production via a heavy new
  resonance $M$, as a function of the resonance mass. Also shown are
  theoretical predictions~\cite{zurek}, assuming a unit coupling. }
\label{fig:limits}
\end{figure}

\begin{figure}[!h]
\subfloat[Singlet models.]{\includegraphics[width=0.75\linewidth]{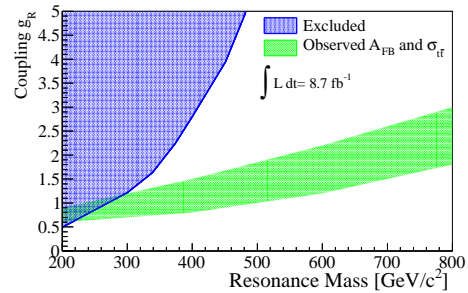}}\\
\subfloat[Triplet
models.]{\includegraphics[width=0.75\linewidth]{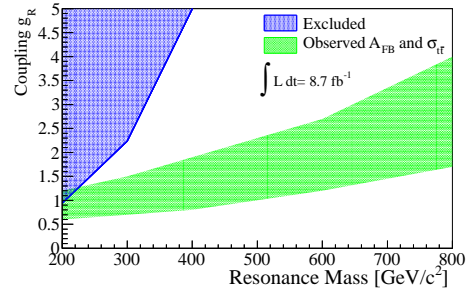}}\\
\caption{Excluded region in the space of resonance mass versus resonance
  coupling ($g_R$) for two specific
  models, where the $M$ particle is part of a new color-singlet (a) or
  color-triplet (b) resonance~\cite{zurek}, respectively. Also shown are
  regions~\cite{goodregion} which are consistent with the  observed
  anomalous $A_{fb}$ and constraints from top-quark pair production
  and single-top production cross-section measurements.}
\label{fig:limits_gm}
\end{figure}

Figure~\ref{fig:sig_region} shows the observed distribution of events
versus $m_{tj}$ 
compared to possible signals and estimated background.  We fit the
most likely value of the sum of the $Mt$
and $M\bar{t} \rightarrow t\bar{t}j$ cross sections by performing a
binned maximum-likelihood fit in the $m_{tj}$ variable, allowing for
systematic and statistical fluctuations via template
morphing~\cite{morph}.  There is no evidence for the presence of top-quark$+$jet resonances
in $t\bar{t}j$ events, so we set upper limits on the combined production
 ($p\bar{p}\rightarrow Mt$ or $\bar{M}\bar{t}$) at 95\% confidence level
 using the CLs method~\cite{MCLimit}.  The observed limits are
 consistent with expectation in the background-only hypothesis
 (Fig.~\ref{fig:limits}). We interpret the observed cross-section
 limit in terms of specific models, one where $M$ is a color singlet
 particle and one
 where $M$ is a colored triplet particle~\cite{zurek}, and construct
 exclusion regions in coupling-mass space~\cite{goodregion}, as shown in Fig.~\ref{fig:limits_gm}. Figure~\ref{fig:limits_gm} also shows that the excluded regions are consistent with the observed anomalous $A_{fb}$
	      and constraints from production cross sections of the top quarks.

In conclusion, we report on the first search for top-quark$+$jet resonances in
$t\bar{t}j$ events. Such resonances are predicted by new physics
models explaining the anomalous top-quark forward-backward production
asymmetry $A_{fb}$. For each accepted event, we reconstruct the 
resonance mass ($m_{tj}$), and find the data to be consistent with SM
background predictions. We calculate 95\% CL upper limits  on
the cross section of such resonance production from $0.61$ pb to $0.02$ pb
 for $M$ masses ranging from $200$ GeV$/c^2$ to $800$ GeV$/c^2$ and interpret the limits in terms of specific
physics models.  These limits constrain a small portion of the
model parameter space. Analysis of collisions at the Large Hadron
Collider may probe the remaining allowed regions.

We thank Kathryn Zurek and Ian-Woo Kim for technical advice and
theoretical guidance. We thank the Fermilab staff and the technical staffs of the participating institutions for their vital contributions. This work was supported by the U.S. Department of Energy and National Science Foundation; the Italian Istituto Nazionale di Fisica Nucleare; the Ministry of Education, Culture, Sports, Science and Technology of Japan; the Natural Sciences and Engineering Research Council of Canada; the National Science Council of the Republic of China; the Swiss National Science Foundation; the A.P. Sloan Foundation; the Bundesministerium f\"ur Bildung und Forschung, Germany; the Korean World Class University Program, the National Research Foundation of Korea; the Science and Technology Facilities Council and the Royal Society, UK; the Institut National de Physique Nucleaire et Physique des Particules/CNRS; the Russian Foundation for Basic Research; the Ministerio de Ciencia e Innovaci\'{o}n, and Programa Consolider-Ingenio 2010, Spain; the Slovak R\&D Agency; and the Academy of Finland.


\begin{thebibliography}{99}
\bibitem{coordinates}
CDF uses a cylindrical coordinate system with
  the $z$ axis along the proton beam axis. Pseudorapidity is $\eta
  \equiv -\ln(\tan (\theta/2))$, where $\theta$ is the polar angle
  relative to the proton beam direction, and $\phi$ is the azimuthal angle  while $p_T = |p|\ \sin\theta$, $E_T = E\ \sin\theta$.
\bibitem{metdef} 
Missing transverse momentum, $\missET$, is defined as the magnitude of the vector $ -\sum_i E_T^i \vec n_i$ where $E_T^i$ are the magnitudes of transverse energy contained in each calorimeter tower $i$, and $\vec n_i$ is the unit vector from the interaction vertex to the tower in the transverse ($x,y$) plane.
\bibitem{goodregion} This region simultaneously satisfies the observed high-$m_{t\bar{t}}$ $A_{fb}$, low-$m_{t\bar{t}}$ $A_{fb}$, and the $t\bar{t}$ cross-section
better than the standard model. Mathematically, it is defined as the region with $\chi^{2}<2.8$, where $\chi^{2}$ is defined in Equation 22 in Ref.~\cite{zurek2}.
$\chi^{2}$ for the standard model is 2.8.
\end{thebibliography}

\end{document}